# The Rise of Quantum Computing – Take a *BITE* for Built Environment and Urban Microclimate Research

**Liangzhu Leon Wang[1,*], Huiheng Liu[1], Honghao Fu[2,*], Zhipeng Deng[3], Bing Dong[4], Naiping Gao[5]**

[1]Centre for Zero Energy Building Studies, Department of Building, Civil and Environmental Engineering, Concordia University, Montreal, Canada

[2]Concordia Institute for Information Systems Engineering, Concordia University, Montreal, Canada

[3]Department of Mechanical and Aerospace Engineering, College of Engineering and Computer Science, University of Central Florida, Orlando, FL, US

[4]Department of Mechanical and Aerospace Engineering, Built Environment Science and Technology (BEST) Laboratory, College of Engineering and Computer Science, Syracuse University, Syracuse, NY, US

[5]School of Mechanical Engineering, Tongji University, Shanghai, China

**Corresponding authors***

Liangzhu Leon Wang, leon.wang@concordia.ca

Professor, P.Eng.

Centre for Zero Energy Building Studies, Department of Building, Civil and Environmental Engineering, Concordia University, Montreal, Canada

https://orcid.org/0000-0002-0653-3612

Honghao Fu, honghao.fu@concordia.ca

Assistant Professor

Concordia Institute for Information Systems Engineering, Concordia University, Montreal, Canada

https://orcid.org/0000-0002-1934-3391

## Abstract

Quantum computing is a new approach to computation that utilizes superposition, entanglement, interference, and tunneling to solve problems too complex for classical computers. This paper discusses the basic concepts and development of quantum computing, exploring its potential applications in the built environment and urban microclimate research. In buildings, quantum computing may help optimize energy management, control HVAC systems, and plan electric vehicle charging networks more efficiently. For urban microclimates, it could accelerate renewable energy planning and support multi-objective design, making it easier to balance urban building performance with climate conditions. Since current quantum hardware is still in the Noisy Intermediate-Scale Quantum (NISQ) stage, we propose the "BITE" principle to guide researchers in choosing suitable problems for quantum acceleration: B (Big search), I (Input-light), T (Tiny computation), and E (Evaluation polish). Although quantum computing still faces challenges such as noise and hardware limits, it offers great potential for developing more climate-resilient, sustainable, and energy-efficient cities of the future.

## 1 Introduction and background

### 1.1 What is quantum computing?

Quantum computing leverages quantum properties like superposition, entanglement, interference, and tunneling to process information in ways impossible for classical systems. The foundations of quantum theory (Dwivedi 2016) can be traced back to Planck's quantum hypothesis (Planck 1901), Einstein's light-quantum proposal (Einstein 1905), and Bohr's atomic model that introduced quantum jumps and stationary states (Bohr 1913). Modern quantum mechanics emerged through matrix mechanics and wave mechanics in 1926(Schrödinger 1926; Born, Heisenberg, and Jordan 1926). These frameworks were unified and extended by Paul A. M. Dirac in 1927 (Dirac 1927), who developed the quantum theory of radiation and the method of second quantization, introduced antimatter through the relativistic Dirac equation in 1928 (Dirac 1928), codified the concept with the Principles of Quantum Mechanics in 1930 (Dirac 1930)and proposed magnetic monopoles to explain charge quantization in 1931 (Dirac 1931). There was a series of debates that led up to the development of Bell's theorem in 1964 (Bell 1964), along with Aspect's experiments in 1982 (Aspect, et al. 1982), which paved the way for quantum information science to exploit nonlocal correlations as a reliable physical resource.

Quantum computing has evolved from a theoretical concept rooted in quantum mechanics into a rapidly advancing field that bridges physics, computer science, and engineering (Hidary 2021). The concept of quantum computation arose from Benioff's quantum Turing machine in 1980 (Benioff 1980), Feynman's quantum simulation in 1982 (Feynman 1982), and Deutsch's universal quantum computer in 1985 (Deutsch 1985). There were two landmark quantum algorithms: Shor's factoring in 1994 (Shor 1994) and Grover's search in 1996 (Grover 1996), which demonstrated super-polynomial and quadratic speedups, respectively, while the introduction of quantum error correction in 1995 (Shor 1995) provided the foundation for fault-tolerant computation. Meanwhile, there were major physical system advances, such as Cirac-Zoller's ion traps (Cirac and Zoller 1995) and Monroe's two-qubit gates in 1995(Monroe et al. 1995), the first Shor NMR (nuclear magnetic resonance) demonstrations (Vandersypen et al. 2001),the superconducting qubits from Cooper-pair boxes in 1999 (Nakamura, Pashkin, and Tsai 1999) to Transmons in 2007 (Koch et al. 2007): the type of superconducting qubits widely used by IBM and Google as the foundation for their quantum processors (Koch et al. 2007). Modern milestones include Google's "quantum supremacy" experiment (Arute et al. 2019), IBM's "pre-fault-tolerant utility" (Kim et al. 2023), and Harvard's neutral-atom logical processor running with encoded qubits in 2024(Bluvstein and others 2024). Together, these breakthroughs are paving the way toward more practical, error-tolerant, and scalable quantum processors. Figure 1 shows the major milestones in the development of quantum theory and quantum computing.

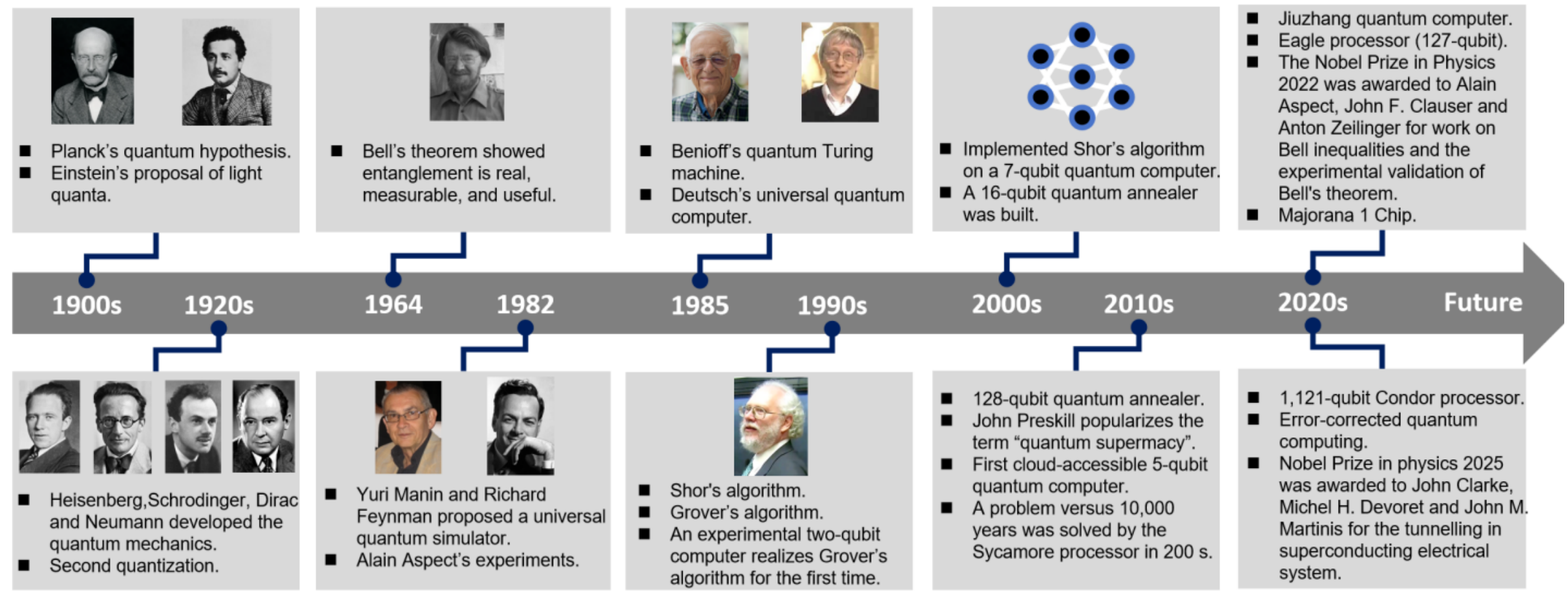

Figure 1. The timeline for Quantum Computing development. (Figure source: Wikimedia Commons, public domain; Bundesarchiv, Bild 183-R57262 / CC BY-SA 3.0 de; Los Alamos National Laboratory; CERN PhotoLab / CC BY 4.0; https://opc.mfo.de/ CC BY-SA 2.0 de; Argonne National Laboratory / CC BY 4.0; https://www.youtube.com/watch?v=8DH2xwIYuT0 /CC BY 3.0; https://www.youtube.com/watch?v=J7HeDX_7Heg&t=7075 /CC BY 3.0))

If we compare a classical computer to a quantum computer, the distinction lies in how information is represented and processed. A classical bit represents either 0 or 1. Even with billions of bits, a modern PC or laptop can only store data as a fixed combination of these two states. In contrast, in quantum computers, data are stored in quantum bits (qubits), which can be in a superposition of being 0 and 1 simultaneously before "measurement". This is analogous to a spinning coin that is both heads and tails while in motion, collapsing into one definite state only upon landing. This means that a superposition of 8 patterns can be represented by 3 qubits at once (Figure 2). Ten qubits can represent 1,024 parallel states, 20 qubits about a million, and 300 qubits can represent more states than there are atoms in the observable universe. We cannot read all that information directly: when we "measure" a quantum state, each qubit still gives us only a 0 or 1, but it shows how much richer the storage is for parallel calculations. The term "measure" or "measurement" is an observation that forces a quantum system to change from a state of multiple possibilities to a single and definite outcome. It is also called "collapse" of the quantum system in the mathematical context of expressing the possibilities as a wave function in quantum theory. To further understand the difference between a modern PC and a quantum computer, imagine a traditional computer as a very fast and reliable light switch: each bit of information is either on (1) or off (0). By flipping the switch on and off, it produces classical information as bit strings. In contrast, a quantum computer is more like a dimmer switch combined with a mirror—it can be on, off, or somewhere in between, and even "reflect" multiple possibilities at once.

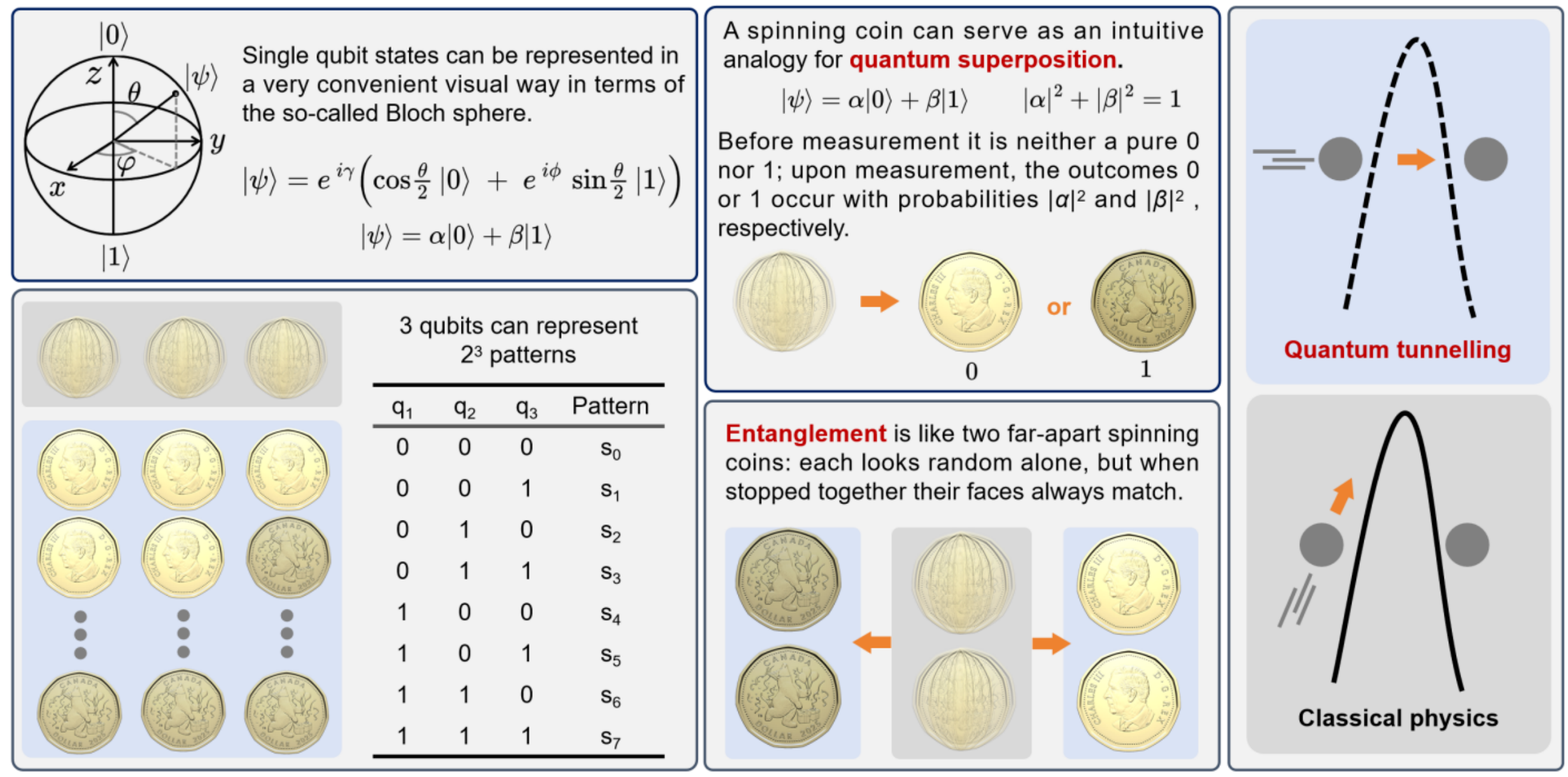

Figure 2. Bloch sphere, coins analogy for quantum superposition, tunneling, and entanglement.

A general qubit is mathematically described by $|\psi\rangle = \alpha|0\rangle + \beta|1\rangle$, where $|0\rangle$ and $|1\rangle$ denote the two possible state vectors in a state space, and the amplitudes $\alpha, \beta$ are complex numbers such that $|\alpha|^2$ and $|\beta|^2$ are the probabilities to measure 0 and 1, respectively (visualized by the Bloch sphere in Figure 2). Note that this requires $|\alpha^2| + |\beta|^2 = 1$. When multiple qubits are connected in a way that their states cannot be described independently, they exhibit a uniquely quantum property known as entanglement. A well-known entangled state between two qubits is the Bell state or EPR pair $\frac{1}{\sqrt{2}}|00\rangle + \frac{1}{\sqrt{2}}|11\rangle$, named after Einstein, Podolsky, and Rosen (EPR) (Einstein, Podolsky, and Rosen 1935). A quantum computer can manipulate many qubits simultaneously, allowing it to explore numerous possibilities at once. Rather than testing each possible solution sequentially, quantum computers leverage interference, a phenomenon similar to the overlapping of wave ripples, to amplify the probability of correct outcomes while canceling out incorrect ones.

The general process of how a quantum computer with many qubits operates can be illustrated conceptually in Figure 2. In the case of many qubits compared to traditional bits, imagine a big table of coins. In a normal PC, the coins (bits) lie flat, either heads (0) or tails (1), and the machine flips them with step-by-step rules (logic gates) to follow one exact pattern. In a quantum computer, the coins are qubits: they can spin in a mix of heads and tails at once (superposition), several coins can be linked (entanglement), and carefully timed nudges (quantum gates arranged in circuits) make some landing patterns reinforced, and others canceled (interference). To solve some optimization problems, some quantum computers also probabilistically let coins slip through barriers (tunnelling) and "roll" over hills and valleys set by an energy function (a "Hamiltonian"). The combination of all these quantum phenomena is what allows for analog processes like quantum annealing or adiabatic quantum computing. The term "Hamiltonian" to describe the "energy" of a quantum system refers to the Hamiltonian operator, corresponding to the total energy (both kinetic and potential energy), and named after the Irish mathematician William Rowan Hamilton, who formulated classical mechanics' dynamics to be described in terms of energy (Hamilton 1835).

Over the years, quantum algorithms (Montanaro 2016) have grown from a handful of brilliant ideas into a powerful and diverse toolkit. It started with algorithms like Shor's in 1994, which could shake up online security by breaking codes, and Grover's in 1996, which offers a better way to search for a needle in a haystack. Since then, the field has blossomed to include quantum walks in 2004 (Ambainis 2004) for analyzing networks and formulas, the Harrow, Hassidim, and Lloyd algorithm (HHL) for crunching massive systems of equations (Harrow, Hassidim, and Lloyd 2009), specialized methods for simulating molecules and materials, and adiabatic optimization for tackling complex planning and scheduling problems. More importantly, quantum computers do not simply guess the answer, but they encode the rules of the problem (checks, cost/Hamiltonian, or physical laws) so interference, entanglement, superposition, and tunneling increase the probability towards finding good answers. However, not all problems benefit from quantum computation, and current devices remain limited by noise and imperfect error correction, which are characteristic of the Noisy Intermediate-Scale Quantum (NISQ) era. Quantum advantage arises primarily when a problem's structure can be translated into quantum circuits or represented as an energy landscape that quantum effects can exploit. For example, Shor's quantum algorithm makes use of interference to solve the integer factorization problem efficiently, a task widely believed to be impossible with classical computers. This is one of the main motivations of this paper to discuss the potential applications of quantum computing to the research of urban microclimate and built environment.

There are three main areas in quantum computing right now. One area is search problems, akin to finding a needle in a haystack. In a normal computer, we must go through all the possibilities until we find the right one, which can be quite time-consuming. The quantum computer can put all the possibilities into play at once and use interference to make it more likely that the right answer will show up. While it does not know the answer, it changes the odds, so we do not have to look through every straw individually. Grover's algorithm achieves a quadratic speedup over classical search algorithms.

A second major area involves simulating molecules, atoms, and materials based on quantum mechanics. While classical computers can only approximate these systems, the amount of data they must handle grows exponentially with system size, quickly becoming intractable. Quantum computers, on the other hand, are built from qubits that obey quantum rules themselves, enabling them to naturally and efficiently model these quantum behaviors. In other words, it could help us understand chemical reactions, design new medicines, or build better batteries and materials. For instance, a quantum simulator is suitable for encoding fluid dynamics transport phenomena within a lattice kinetic formalism (Mezzacapo et al. 2015). Quantum algorithms for the wave equation can act as a fast, memory-efficient kernel to accelerate wave-dominated modules in CFD (Costa, Jordan, and Ostrander 2019). Quantum lattice Boltzmann method can be plugged into CFD pipelines as an accelerated time-stepper for incompressible flows (Budinski 2022).

Another major area is combinatorial optimization, a process of finding the best arrangement among many options. Examples include finding the best shipping/travel route, the most efficient schedule, or the lowest-energy configuration of a system. Classical computers usually explore the problem step by step, which risks getting stuck in a "good enough" solution without finding the optimal one. This is sometimes known as a "local" or "false" solution. In comparison, quantum effects, such as tunneling and entanglement, enable the system to slip through barriers among local optima and explore deeper valleys in the energy landscape. This makes the best or near-best answers more likely to appear when we measure the qubits. The process for this task is often called "adiabatic quantum computing" (Farhi et al. 2001) or "quantum annealing". The word "annealing" comes from materials science, describing a heat treatment technique for changing the physical properties of a metal or glass through heating and cooling, allowing the materials to relieve internal stresses and reach a final state of the lowest possible energy and thus high stability. Quantum annealing mirrors this physical process in a computational context. To illustrate, imagine a cloud drifting over a foggy mountain range, capable of overseeing the whole range while seeking the lowest valleys. Through superposition, it spreads across many parts of the landscape at once, sensing multiple valleys simultaneously. And thanks to quantum tunneling, portions of the cloud can slip through thin ridges rather than having to climb over them, allowing it to flow more easily into deeper valleys and settle into the lowest one more efficiently. In contrast, a classical computer explores this landscape point by point, akin to rolling a ball down the mountain, which can be easily trapped in small dips. Although quantum annealing does not guarantee the best answer every time, it offers a promising shortcut for certain types of problems. Quantum annealing shines in solving combinatorial optimization problems like spin-glass and Ising models (Lucas 2014), Traveling Salesman Problem, Maximum Independent Set, Job-Shop Scheduling (Hauke et al. 2020), and protein-folding energy minimization (Durant et al., 2024).

Besides quantum annealing, variational quantum algorithms (VQAs) are a prominent family of hybrid quantum–classical methods, and VQAs use a parameterized quantum circuit (often called an ansatz) whose parameters are adjusted by a classical optimizer to minimize an objective function (Cerezo et al. 2021). This ansatz is then trained in a hybrid quantum-classical loop to solve the optimization task. The Quantum Approximate Optimization Algorithm (QAOA) (Farhi, Goldstone, and Gutmann 2014) is a canonical VQA for combinatorial optimization, which is a quantum operation depending on a set of continuous or discrete parameters that can be optimized. The process can be evaluated quickly on a quantum computer. A lower value of the objective function corresponds to a better solution. By adjusting the parameters, we can reliably push the value down until it reaches that lowest point. Furthermore, quantum machine learning (QML) offers a suite of potential applications for small quantum computers (Biamonte et al. 2017). These quantum algorithms can be applied to the built environment and urban microclimate to speed up the equations' solution process based on quantum-gated methods.

Engineering optimization problems have been explored on several quantum platforms. Quantum annealers (e.g., D-Wave) have been used in pilot studies for transportation routing, such as demonstrations to improve bus routing under certain constraints (Kulawik, D-Wave, and Andrade 2019). These should be described as proof-of-concept or pilot demonstrations rather than conclusive production deployments. Gate-based and hybrid quantum approaches (QAOA and other VQAs) are being studied for grid management, resource scheduling, and supply-chain optimization. Industry research groups such as Siemens are investigating quantum-inspired and hybrid optimization methods for resilience and loss reduction (Singh, 2024). These pilot applications show that quantum algorithms, quantum-inspired optimization, and hybrid quantum–classical approaches are a promising direction for engineering optimization; however, important challenges remain (hardware scale, noise, embedding overheads, and optimizer tuning). For engineering optimization tasks, quantum computing platforms differ significantly in their hardware architecture, programming models, and maturity for production workflows. Quantum annealers, such as D-Wave systems (D-Wave 2025), are purpose-built to solve combinatorial optimization problems, such as routing, scheduling, and energy grid balancing. Instead of general-purpose algorithms, they are designed for engineers tackling problems that can be written as Quadratic Unconstrained Binary Optimization (QUBO) (Glover et al. 2022) with (0 and 1), closely related and convertible to the Ising models, which describes the energy of a system of magnetic spins in either "up" (+1) or "down" (-1).

Cloud environments such as Amazon Braket, D-Wave Leap, and Microsoft Azure Quantum provide multi-vendor access and orchestration: users can experiment with different backends (annealers, trapped-ion, superconducting gate devices, neutral-atom machines) and compare algorithmic performance in a single ecosystem. Gate-model hardware (IonQ, Quantinuum, Rigetti, IBM, QuEra, Pasqal, etc.) supports algorithms such as the QAOA and variational approaches, but NISQ-era limitations (noise, constrained circuit depth, and limited qubit counts) currently restrict their advantage to carefully selected problem instances or to algorithmic research. In short, the choice of provider and hardware class should be driven by: (1) how naturally the problem maps to QUBO/Ising; (2) the embedding overhead and available hybrid solvers; (3) reproducibility and cost; and (4) whether performance metrics (objective gap, time-to-solution, and scalability) surpass classical baselines for the specific problem class. In terms of qubit count offered from these suppliers, as of 2025, D-Wave offers over 4400 qubits, which are specifically designed for optimizations and are different from the general ones of Google (54), IBM (~1121), Microsoft (8), IonQ (36), Quantinuum (56), Rigetti (36), and QuEra (256).

In the following section, we focus on how quantum computing can solve optimization problems in the built environment and urban microclimate. We will talk about its potential, applicability, and limitations as it pertains to urban microclimates and the built environment.

## 1.2 Built environment and urban microclimate engineering research

Built environment is all the surroundings that humans create or modify to facilitate human activity, including buildings, infrastructure like transportation networks, public spaces, and urban landscapes. Most key research areas deal with complex, interrelated issues in the built environment, such as energy consumption, economic costs, carbon emissions, and occupant comfort (Chen, Tsay, and Ni 2022). Building simulations are widely used for building energy analysis, thermal resilience performance assessments, thermal comfort and air quality assessments, life cycle assessments (LCAs), integrating renewable energy systems into buildings, optimizing building designs, predicting occupant behavior in buildings, and managing traffic flows. There are three types of building simulations: white box, based on fundamental physical principles (like building energy models, such as EnergyPlus), gray box, which uses simplified equations with thermal resistance and capacity, historical data integrations, and black box, based on data-driven algorithms like artificial neural networks and long-short term memory networks (de Wilde 2023). When we compare different types of building simulations: physics-based model wins for what-if design, grey-box wins for calibrated control, and black-box excels for speed/scale. The current major trend of building simulations, especially building energy simulations, moves from single-building applications to urban scales, such as urban building energy modeling (UBEM) studies (Salvalai, Zhu, and Sesana 2024; Ali et al. 2021), which emphasize uncertainty treatment, archetype robustness, and systematic calibration. In terms of building information studies, BIM (Ahmed, Heywood, and Holzer 2025) helps with decision-making processes like cost analysis, carbon footprint management, and life cycle assessment.

An urban microclimate is a distinct local atmosphere within a city that differs from the surrounding countryside, mainly due to buildings, paved surfaces, and humans. Urban Heat Island (UHI) altered wind patterns, variations in precipitation, and air quality fluctuations are some of the key research topics (Yang et al. 2023). Due to buildings and pavements absorbing heat, human activity releasing waste heat, and reduced vegetation, cities usually have higher temperatures than their rural surroundings. In addition to obstructing and channeling wind, buildings also contribute to air pollution and degrading air quality. In urban microclimatology, experimental approaches like field measurements and wind tunnel simulation, computational fluid dynamics simulations based on Navier-Stokes equations, and data-driven methods like multiple linear regression, nonlinear regression, random forests, AI, recurrent neural networks, and neural operators, such as Fourier neural operators(Peng et al. 2024; Qin et al. 2025), are used. The current major limitations of using AI/MLs include the lack of high-quality urban data, the generalization of specific black-box models (Tam et al. 2022), and complex boundary conditions and their inclusion in the models. To mitigate ML limitations, researchers are exploring physics-informed ML, hybrid physics-ML surrogates, and transfer learning across cities.

# 2 How to apply quantum computing?

## 2.1 Quantum computing for built environment

Quantum computing can advance built environment research and application by solving complex optimization, optimal control, and simulation problems for building design and operation. This involves structural optimization, exploration of novel sustainable materials, optimization of the construction flow scheme, and the design of more efficient and resilient energy and transport network systems. In optimal control, for instance, the on/off status of chillers can be encoded as binary variables to derive an energy-cost minimization objective function in hours. Load demand-supply balance and minimum runtime constraints can further be employed to support day-ahead scheduling of building Heating, Ventilation, and Air Conditioning (HVAC) systems. For thermal resilience assessment and demand response, charging or discharging of thermal energy storage can be represented by binary variables, enabling objectives that minimize peak grid load and thermal discomfort. Additional constraints can be energy balance and mutual exclusivity between charging and discharging, which allow assessment of system flexibility. In intelligent building design and construction, binary variables can represent the open/closed status of windows and whether the curtain is drawn, supporting objectives that balance HVAC energy use and thermal comfort under budget constraints, thereby guiding adaptive smart-window strategies.

On the grid-interactive building side and urban scale, real-time demand response involves thousands of devices, uncertain weather, and occupancy (Han et al. 2025; Deng, Wang, and Dong 2025). While stochastic comfort models account for parameter uncertainty, they still face major challenges related to the intermittency of renewable sources and the resulting explosion of scenarios, which hinder reliable and scalable control. The combination of discrete options, tight timing, and uncertainty makes grid-interactive building coordination a tough class of optimization problems. EV charging infrastructure introduces more coupling, where to site chargers, how many sites, what types to install, and when to charge fleets (Hammam, Nayel, and Mohamed 2024). When city-scale locations are coupled with time-dependent operations, they yield large, sensitivity-prone models under uncertain travel demand as EV adoption grows (Deng et al. 2025). The interplay of network constraints, time-of-use economies, and mobility patterns is why these issues are both high-value and high-complexity.

## 2.2 Quantum computing for urban microclimate

Quantum computing is promising for boosting the speed and quality of urban renewables and informatics, climate modeling, such as high-fidelity simulation of airflow, heat transfer, and pollutant transport in cities. For example, current research focused on quantum CFD (Jaksch et al. 2023) and quantum physics-informed neural networks (PINN) (Sedykh et al. 2024). However, directly solving the continuous or discrete Navier–Stokes equations on near-term quantum hardware is not considered feasible. Instead, quantum methods are more realistically applied to discretized or reduced-order CFD representations, in which flow variables (e.g., velocity, pressure, or pollutant concentration) are defined on a lattice or coarse mesh. Under steady-state or surrogate modeling assumptions, these problems can be expressed as the minimization of an energy functional subject to discretized conservation constraints. After discretization, such functionals can be mapped onto lattice-based Hamiltonians in the Ising or QUBO form, where binary variables encode discretized states, regime selections, or boundary-condition configurations. In this hybrid framework, the quantum solver functions as a combinatorial optimizer, that explores the vast, non-convex landscape of the problem to identify high-potential candidate solutions. By navigating these rugged energy terrains, the quantum hardware provides a set of optimized configurations or shortlisted states that would be computationally expensive for classical solvers to find. These results then serve as an intelligent starting point, which is subsequently refined and validated using high-fidelity classical simulations to ensure physical accuracy and convergence.

At the city scale, for example, when deploying solar photovoltaic (PV) systems across a large urban area, binary variables representing installation status and panel tilt angle can be introduced to formulate a multi-objective function that considers power generation, public budget constraints, and carbon emissions. Stepping up to UBEM, even the seemingly simple task of calibrating a city-scale model, e.g., through adjustments to archetypes, schedules, and systems, poses significant challenges due to substantial input uncertainty and sparse observational data. After being adjusted for UBEM, which often becomes an input for the very optimization tasks, such as urban retrofits, district heating and cooling arrangements, grid-interactive efficient building operating sequences, error and uncertainty propagate. These challenges motivate the need for optimization frameworks that explicitly handle uncertainty and large discrete decision spaces. In this context, UBEM calibration can be formulated as a binary optimization problem by discretizing uncertain parameters into finite candidate sets and encoding them with binary variables. The calibration objective minimizes discrepancies between simulated and measured energy use across many buildings, while penalty terms enforce physical feasibility and policy constraints. This formulation naturally leads to a QUBO or Ising representation, where quantum or hybrid quantum–classical solvers can accelerate the search over high-dimensional discrete parameter spaces, rather than replacing the underlying physics-based simulation. Quantum computing can tackle the complex interplay between the built environment and urban microclimate by allowing integrated modeling and multi-objective optimization. It could enable holistic design strategies where building performance and urban planning strategies are optimized together in terms of energy efficiency, occupant comfort and beneficial microclimatic effects towards advancing resilient and sustainable urban development. Consider a mid-rise neighborhood consisting of ten buildings surrounded by an area of plaza. A quantum optimization framework integrates detailed building energy models that account for HVAC loads, envelope characteristics, and occupancy-driven internal gains. For each building, facade orientation, glazing ratio, and thermal insulation characteristics can be modeled as continuous optimization variables, where plaza features such as tree locations, pavilion albedo, and water feature locations can be modeled as binary variables. Then a multi-objective cost function may minimize annual energy use, peak summer plaza temperature, and mean indoor discomfort hours altogether, assuming objective functions can be identified. By embedding this in a QUBO Hamiltonian, a quantum annealer searches for the enormous design space in superposition without the need for gradient computations and may rapidly converge to Pareto-optimal solutions that make possibly the best trade-offs.

There are other examples: Heat-mitigation options (cool pavements of roofs, greenery) have robust evidence bases to use the binary choices. Specifically, to enhance thermal comfort and provide co-benefits for PV performance by lowering urban and roof-skin temperatures by several degrees, the use of reflective "cool" roof/wall materials and green roofs represents discrete design choices (Santamouris 2014), which can therefore be modeled as binary variables. These give measurable deltas that can be encoded in the objective and constraints (cost, feasibility, maintenance).

In summary, research on the built environment and urban microclimates often involves systems comprising many uncertain parameters and requiring combinatorial validation across numerous target optimal conditions. Such complexity pushes the limits of classical computing in capturing full search domains with increasingly intricate boundary conditions. Classical large-scale optimization algorithms often require exponential time, but quantum annealing (QA) can provide speedup through quantum parallelism, as well as optimizing over a vast state space, e.g., $2^{300}$ states for 300 qubits. Superposition, as well as entanglement, enable the exploration of many solutions by means of QA at once, as well as tunneling over sizable high-energy barriers with rough landscapes.

# 3 When is quantum computing appropriate or not?

**Some necessary conditions/considerations for quantum computing:**

There is no single rule that determines whether a problem will benefit from quantum computing; however, a set of pragmatic criteria reliably identifies suitable cases. A problem is promising for current quantum computers if it can admit a compact and hardware-compatible encoding (binary/Ising/QUBO) and has the kind of "rough" search landscape where quantum effects help. Rugged landscapes with many local traps, narrow barriers, and nonlinear and nonconvex problems are more promising than smooth, convex problems (Albash and Lidar 2018). With the need for fast computation and real-time simulation requirements for complex urban-scale situations, quantum computing is necessary, especially for accelerating certain Nondeterministic Polynomial-Time (NP) problems. If we need lots of extra variables or very deep "circuits", the potential benefits may vanish. There are also hardware limits in the NISQ (Noisy Intermediate-Scale Quantum) era: the number of qubits, how they connect, noise, and the cost of embedding or routing. These can quickly turn a feasible model into an impractical one. The **objective and constraints must be fast to evaluate** (ms/s per call) because the hybrid quantum–classical loops call them many times. It is always preferred to pick problems that are **big enough to matter**, but not so big that we cannot encode them, and avoid workflows where **data loading dominates**.

**Possible sufficient conditions when quantum computing is not needed:**

By contrast, quantum computing is generally not advantageous for problems that are near-convex and efficiently solved by established classical optimization paradigms (Linear Programming/Semi-definite Programming) with strong guarantees (Afram and Janabi-Sharifi 2014), or if full-fledged classical methods already reach our targets. If quantum mapping blows up variables or couplings, or hardware routing necessitates deep circuits, the problem is not suitable for quantum computing (Sporleder, et al. 2022).

QC and classical ML/AI generally target distinct computational tasks. Quantum annealing and QAOA mainly tackle discrete combinatorial optimization through QUBO/Ising formulations; for many families of NP problems, worst-case scaling remains exponential for both quantum and classical solvers. Therefore, any claimed "advantage" must be practically tested. This evaluation should compare the time needed to find a solution and the quality or feasibility of that solution against strong classical methods, such as Mixed-Integer Linear Programming, Genetic Algorithms, and Simulated Annealing. In comparison, classical ML/AI is widely used for prediction and creating surrogate models. Here, the challenge of training grows with the amount of data and the number of model parameters. This field benefits greatly from established and efficient GPU and TPU computing systems. Consequently, in the near term, quantum computing is more likely to serve as a complementary tool. Its potential uses include optimizing discrete decisions or speeding up specific numerical tasks. It is not positioned to replace mainstream machine learning training. Formal quantum advantage is confirmed when a quantum system outperforms the best-known classical method. This superior performance is demonstrated by a shorter Time-to-Solution (TTS), higher-quality results, or better approximation accuracy. The quantum system must also show more favorable scaling as problems become larger. All this must be achieved while remaining robust against real-world hardware noise and decoherence.

In practical workflows, therefore, we recommend choosing problems that are large in combinatorial choice space but small in per-evaluation data footprint: encode sparsely, reduce per-call complexity potentially with surrogates (Forrester and Keane 2009), if possible, or reduced-order models, warm-start the quantum solver (Egger, Marecek, and Woerner 2021), and use quantum methods primarily as a high-quality diversifier whose output is polished by a classical refinement. This hybrid strategy balances the strengths and current limitations of near-term quantum.

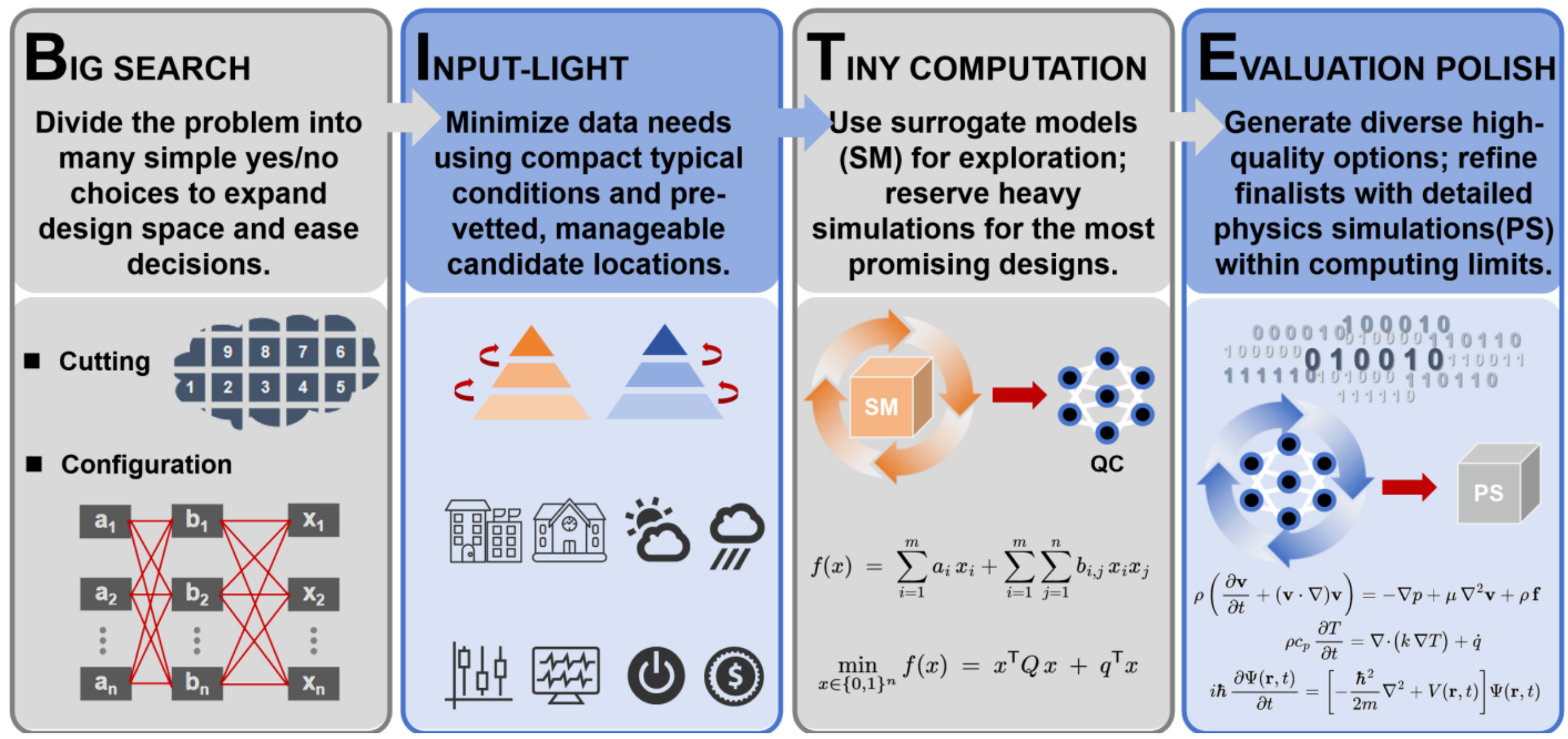

Figure 3. The "BITE" rule for the application of Quantum Computing

As a practical guide for problem identification and structuring, follow the **BITE** rule in Figure 3: **Big search, Input-light, Tiny computation (inner step), Evaluation polish (outer step).**

- **B – Big search**: Break large, complex problems into many small and simple decisions. For example, an entire city can be divided into a set of small spatial tiles. This approach creates a vast design space without making individual decisions overly difficult. Framing challenges as numerous yes-or-no binary choices is often more effective than relying on a few highly interdependent variables that demand intensive modeling.
- **I – Input-light**: Keep data requirements to a minimum so that the optimization problem can be easily uploaded to a quantum computer. Instead of generating massive new datasets, rely on smaller, well-structured data sources that can be easily managed and updated.
- **T – Tiny computation**: In the central phase of the workflow, use fast and lightweight models such as surrogate models derived from large-scale models so that the quantum computer can quickly estimate temperature, airflow, or energy outcomes. Reserve detailed, computationally heavy simulations (e.g., fully-dynamic or high-resolution energy models) for only the most promising design candidates for the next classical postprocessing step. This balance enables rapid exploration and optimization while keeping the quantum computational demands practical.

- **E – Evaluation polish**: The main role of quantum tools is to generate a broad set of diverse, high-potential solutions. After this exploration, refine the top candidates using detailed, physics-based simulations on classical computers. This step verifies accuracy and ensures that the process remains within the available computational budget.

A simple guiding mantra: *"Let quantum roam a **Big** space of choices, keep the **Input** light, make each quantum **inner-step Tiny** with fast surrogates, and save heavy physics for **Evaluation polish in the classical postprocessing**."*

Here is a q*uick* sanity check for the multi-level setups:

- **Big?** - Many sites, each with a few discrete choices, not a continuous space**.**
- **Input-light?** - Compact layers and reduced scenarios, no heavy data ingestion.
- **Tiny (inner loop)?** - A quadratic surrogate over level encodings; millisecond scoring.
- **Evaluation (outer loop)?** - Full-physics simulations only for finalists, with optional active learning to refine the surrogate where it performs poorly.

# 4 Conclusions

Quantum computing, rooted in quantum mechanics, is gradually moving from theory to practical application. To develop further, it could be freed from classical limitations, especially for difficult systems and applications in the built environment and urban microclimates (whereby computing systems become less and less practical). Thus, this perspective paper introduces the basic concepts, latest development trends, and applications of quantum computing in buildings and urban science, including its applications and advantages in complex optimization, system simulation, and integrated modeling. Our findings suggested that quantum computing can be beneficial for specific problems in building energy management and urban climate scenarios. An illustrative case of this is real-time HVAC control, scheduling electric vehicle charging, or optimizing a region-wide energy network, which can effectively be defined in quadratic unconstrained binary optimization (QUBO) problems (quantum annealing or variational quantum algorithms can easily be applied). Quantum computing also supports urban-scale simulations, including EV charging and UBEM, and advances physics-informed neural networks (PINNs) to simulate urban settings at high resolution.

We need to keep in mind that quantum technology is still in its infancy. Hardware limitations (such as noise, small number, and low-quality qubits), imperfect error correction processes, in addition to the adaptation of real-world to quantum models, are major hurdles. Quantum computation will not make its way to every problem. Quantum techniques have a marginally positive impact on near-convex problems, or the ones that can be solved quite nicely with the aid of efficient classical methods.

To help us discover suitable applications, we introduce the "BITE" guideline—a stepwise guide for choosing and organizing a series of problems with promising opportunities for quantum acceleration. It highlights four properties: B (Big search), I (Input-light), T (Tiny computation), and E (Evaluation polish).

In the future, quantum computing is expected to play a powerful role in building and urban science sectors. Future research studies should focus on quantum computing, customized problem formulations, optimal quantum-classical hybrid workflows, and joint work between physical science, computing/building/urban science, and engineering. Quantum computing in the complex systems optimization context is also expected to emerge as an important innovation that addresses problems of urban sustainability and allows for more climate-resilient, energy-efficient, and occupant-centric livable communities and cities, when quantum systems hardware and methods are becoming more mature, powerful, and optimized.

## Acknowledgement

We acknowledge the reviews, discussions, and inputs from Dr. Ken Robbins from D-Wave Quantum, and the financial supports from the Natural Sciences and Engineering Research Council (NSERC) of Canada through the Discovery Grants Program [#RGPIN- 2024-06297], and the Canada First Research Excellence Fund (CFREF) [IMPACT Project - Transforming Built and Urban microclimates: Advancing Resilience Science for Vulnerable Populations in a Decarbonized and Electrified Canada].